\documentclass[pra,aps,showpacs,twocolumn,floatfix,preprintnumbers]{revtex4}
\usepackage[]{graphicx}
\usepackage{dcolumn,epsfig}
\usepackage{amsmath}
\usepackage{hyperref}
\usepackage{color}

\begin{document}

\title{Titania-doped tantala/silica coatings for gravitational-wave detection}

\author{Gregory~M~Harry,  Matthew~R~Abernathy, Andres~E~Becerra-Toledo}
\address{LIGO Laboratory,  Massachusetts Institute of Technology,
\\ NW17-161, Cambridge MA, 01239, USA}
\email{gharry@ligo.mit.edu}

\author{Helena Armandula, Eric Black, Kate Dooley, Matt Eichenfield,
Chinyere Nwabugwu, Akira Villar}
\address{LIGO Laboratory, California Institute of Technology,
\\ Pasadena CA, 91125, USA}

\author{D~R~M~Crooks, Gianpietro~Cagnoli, Jim~Hough, Colin~R~How, Ian~MacLaren,
Peter~Murray, Stuart~Reid, Sheila~Rowan, Peter~H~Sneddon}
\address{Department of Physics and Astronomy, The University of
Glasgow, Glasgow, G12 8QQ, United Kingdom}

\author{Martin~M~Fejer, Roger~Route}
\address{Edward L Ginzton Laboratory, Stanford University, Stanford
CA, 94305, USA}

\author{Steven~D~Penn}
\address{Physics Department, Hobart and William Smith Colleges,
Geneva NY 14456}

\author{Patrick Ganau, Jean-Marie Mackowski, Christophe Michel, Laurent Pinard, Alban Remillieux}
\address{Laboratoire Mat{\'e}rial Advanc{\'e}e-Virgo, Lyon, France}

\begin{abstract}Reducing thermal noise from optical coatings is crucial to
reaching the required sensitivity in next generation interferometric
gravitational-waves detectors.  Here we show that adding TiO$_2$ to
Ta$_2$O$_5$ in Ta$_2$O$_5$/SiO$_2$ coatings reduces the internal
friction and in addition present data confirming it reduces thermal
noise. We also show that TiO$_2$-doped Ta$_2$O$_5$/SiO$_2$ coatings
are close to satisfying the optical absorption requirements of
second generation gravitational-wave detectors.
\end{abstract}

\pacs{04.80.Nn,62.40.+i, 95.55.Ym}

\maketitle 

\section{Introduction}
\label{sect:intro}  

Interferometric gravitational-wave detectors are now operating in
the United States\cite{S1Instrument},
Europe\cite{S1Instrument,virgo}, and Japan\cite{tama}.  Second
generation detectors\cite{advLIGO} are being designed which will
have the sensitivity to make likely an actual detection of a
gravitational wave\cite{Thorne_Cutler}.  This will require reducing
all noise sources, but especially the thermal noise which is
predicted to be the limiting noise in the most sensitive band around
100~Hz. Much of this thermal noise will come from the optical
coatings of the interferometer mirrors\cite{syracuse}.  Coatings
with improved thermal noise performance will allow for greater
sensitivity to gravitational waves and improved astrophysical
performance. Coating thermal noise is also the limiting noise source
for laser frequency stabilization~\cite{Jordan}, making this
research effort important for other precision experiments.

Thermal noise is caused by mechanical loss in the system in
accordance with the Fluctuation-Dissipation Theorem\cite{Callen}.
Directly applying this theorem to the case of a Gaussian-profile
laser sensing the position of a coated mirror yields\cite{Yuri}
\begin{equation}
\label{eqn:tn} S_{x}\left(f\right) = 2 k_B T
\phi_{\mathrm{eff}}\left(1-\sigma^2\right)/\left( \pi^{3/2} f w
Y\right),
\end{equation}
for the thermal noise.  Here $S_x\left(f\right)$ is the power
spectral density of position noise, $k_B$ is Boltzmann's constant,
$T$ is the temperature, $\sigma$ is the Poisson ratio of the
substrate material, $w$ is the half-width of the Gaussian laser
beam, and $\phi_{\mathrm{eff}}$ is the effective loss angle of the
mirror.  The loss angle can be written as\cite{Ottawa}
\begin{eqnarray}
\label{eq:phi} \phi_{\mathrm{eff}} &=& \phi +
d/\left( \sqrt{\pi} w Y_{\perp}\right) \\
& & \left( \left( \left( Y/\left(1-\sigma^{2}\right) - 2
\sigma_{\perp}^2 Y Y_{||}/\left(Y_{\perp}
\left(1-\sigma^2\right)\left(1-\sigma_{||}\right)\right)\right)
\phi_{\perp} \right. \right. \nonumber \\
& & \left. + Y_{||} \sigma_{\perp} \left(1-2
\sigma\right)/\left(\left(1-\sigma_{||}\right)\left(1-\sigma\right)\right)\left(\phi_{||}
- \phi_{\perp}\right) \right. \nonumber \\
& & \left. + Y_{||} Y_{\perp} \left(1+\sigma\right)\left(1-2
\sigma\right)^2/\left(Y\left(1-\sigma_{||}^2\right)\left(1-\sigma\right)\right)\phi_{||}\right),\nonumber
\end{eqnarray}
where $d$ is the coating thickness, $Y$, $\sigma$, and $\phi$ are
the Young's moduli, Poisson's ratios, and loss angles of the
substrate (no subscript), and for the coating for stresses
perpendicular ($_{\perp}$) and parallel ($_{||}$) to the optic face.

In addition to low thermal noise, the coatings must also satisfy
strict thermal and optical requirements.  The Fabry-Perot cavities
that make up the arms of these detectors must have high finesse,
limiting the acceptable transmission and scatter to a few parts per
million (ppm).  In addition, the transmission must be matched
between mirrors to better than 1 percent so that the two arms will
have nearly equal finesse.  The absorption has a stricter
requirement, better than 0.5 ppm, due to thermal lensing
considerations\cite{Ryan}. Thermal considerations also dictate that
the absorption be as uniform as possible across the face of the
optic.

\section{Background}
\label{sect:background}

Study of multilayer dielectric optical coatings is an ongoing
research project in the gravitational-wave detection
community\cite{syracuse,Ottawa,Yamamoto1,Yamamoto2,Beauville,Crooks,Penn,Crooks2,tni,Fejer,Braginsky03}.
The coating used in initial interferometers, alternating $\lambda/4$
layers of SiO$_2$ and Ta$_2$O$_5$, was studied to determine if the
mechanical loss was enough to cause thermal noise
problems\cite{syracuse,Crooks}. The particular coatings measured
were coated by Research-Electro Optics (REO) of Boulder CO, USA.
When it was determined that the loss was enough to cause limiting
noise in next generation gravitational-wave detectors, research was
carried out to determine the source of the mechanical loss in
SiO$_2$/Ta$_2$O$_5$ coatings\cite{Penn}. This was done in
collaboration with LMA/Virgo of Lyon, France, who coated the
samples. This established that the loss came from internal friction
in the coating materials rather than any interface effects between
layers or between the substrate and the coating.  It was also found
that the Ta$_2$O$_5$, rather than the SiO$_2$, was the dominant
contributor to the coating mechanical loss.  The loss angles of the
 SiO$_2$ and the Ta$_2$O$_5$ were found to
be\cite{Crooks2}
\begin{eqnarray}
\label{eq:phisio2} \phi_{\mathrm{SiO2}} & = & \left( 1.0 \pm 0.2
\right) \times 10^{-4} + f \left( 1.1 \pm 0.5 \right) \times
10^{-9},
\end{eqnarray}
\begin{eqnarray}
\label{eq:phita2o5} \phi_{\mathrm{Ta2O5}} & = & \left( 3.8 \pm 0.2
\right) \times 10^{-4} + f \left( 1.8 \pm 0.5\right) \times 10^{-9},
\end{eqnarray}
as a function of frequency $f$.

The thermal noise from optical coatings has also been directly
observed in two small scale interferometers, one at
Caltech\cite{tni} and one in Japan\cite{Numata}. The Caltech
measurement was on an REO coating and found thermal noise at a few
kiloHertz to be consistent with
\begin{equation}
\phi_{\mathrm{eff}}  = 6.5 \pm 0.4 \times 10^{-6}, \label{eq:tni}
\end{equation}
where $\phi_{\mathrm{eff}}$ is from Eq.~\ref{eq:phi}. The values
determined from the modal Q measurements in Eqs.~\ref{eq:phisio2}
and \ref{eq:phita2o5} predict
\begin{equation}
\phi_{\mathrm{eff}}  = 6.4 \pm 0.3 \times 10^{-6}.
\end{equation}

Optical absorption was also measured for Ta$_2$O$_5$/SiO$_2$
coatings. Coatings from REO and LMA/Virgo gave similar
results\cite{Cimma},
\begin{eqnarray}
\alpha_{\mathrm{REO}} & = & 0.3 \pm 0.1~\mathrm{ppm} \\
\alpha_{\mathrm{LMA}} & = & 0.4 \pm 0.1~\mathrm{ppm}.
\end{eqnarray}

\section{Measurement}
\label{sect:measurement}

The next stage of coating research has been to improve the
mechanical loss without significantly degrading the optical
absorption.  Adding TiO$_2$ as a dopant to the Ta$_2$O$_5$ was tried
because it has a high Young's modulus, its atomic size allows for
dense packing in the Ta and O matrix, and the melting point of
TiO$_2$/Ta$_2$O$_5$ alloy is relatively high, indicating a stable
amorphous structure.

Silica substrates were coated with the TiO$_2$-doped
Ta$_2$O$_5$/SiO$_2$ coating using ion beam deposition. Details of
the coating process can be found in a recent paper~\cite{jmm}. There
were two different coating chambers used to make the samples
studied, one large and one small.  The primary difference between
the chambers is that the ion source in the small chamber is a
Kaufman source with tungsten filaments, whereas in the large chamber
it is two Radio Frequency ion sources. In the small chamber, the
tungsten filament heats the target as well as the substrate.  After
coating, each sample was annealed at 600 degrees C.  X-ray
examination showed that no large crystals had formed in the coating
after annealing.

Each coating consisted of 30 $\lambda/4$ (at 1.064~$\mu$m) layers
alternating between the two materials, TiO$_2$-doped Ta$_2$O$_5$ and
SiO$_2$.  The total thickness was measured two ways, using
reflectivity measurements and with an electron microscope.  The
methods agreed with each other within 5\%, for an average coating
thickness of $4.5 \pm 0.1 \mu$m. One sample was different, a single
layer of TiO$_2$-doped Ta$_2$O$_5$ 4.7~$\mu$m thick.

The concentration of TiO$_2$ in Ta$_2$O$_5$ for each coating was
measured two different ways.  First, an estimate was made by
comparing the index of refraction of the TiO$_2$-doped Ta$_2$O$_5$
with pure Ta$_2$O$_5$ and pure TiO$_2$.  A linear relationship was
assumed between TiO$_2$ concentration and index so the TiO$_2$
concentration was obtained by interpolation.  This is only valid
when done between coatings from the same chamber, large or small. A
more detailed measurement was made on some samples using electron
energy loss spectroscopy, which is described in the Appendix. The
two methods agreed fairly well when the same coating was studied by
both, as seen in Table~\ref{table:conc}.

\begin{table} \caption{Concentration of TiO$_2$ in Ta$_2$O$_5$
as measured by change in index of refraction and by electron energy
loss spectroscopy (EELS).} \label{table:conc}
\begin{center}
\begin{tabular}{lrr}
\hline
Coating & [TiO$_2$] - Index & [TiO$_2$] - EELS \\
\hline
0 & 0\%  & -  \\
1 & $6 \pm 0.6$ \%  & $8.5 \pm 1.2$  \\
2 & $13 \pm 1$ \% & -  \\
3 & $24 \pm 2$ \%  & $22.5 \pm 2.9$  \\
4$^{*}$ & $54.5 \pm 5$ \% & $54 \pm 5$  \\
5$^{*}$ & $14.5 \pm 1$ \% & -    \\
6$^{**}$ & $6 \pm 0.6$ \% & -  \\
\hline \multicolumn{3}{l}{$^{*}$ Coated in large coating chamber} \\
\multicolumn{3}{l}{$^{**}$ Single layer of TiO$_2$-doped
Ta$_2$O$_5$}
\end{tabular}
\end{center}
\end{table}

\subsection{Mechanical loss}

Coated silica disks were used to determine the mechanical loss in
the coating.  All disks were 7.6~cm in diameter, with some 2.5~cm
thick and some 0.25~cm thick.  The thicker disks were suspended in a
wire sling, and had normal mode Q's measured with an interferometric
readout.  The thinner disks were suspended with a welded silica
suspension and had modal Q's measured using a birefringence readout.
Details of the suspension and readout systems for both types of
disks can be found in a recent publication\cite{Penn}.

Modal Q's were measured on multiple modes of both thin and thick
samples for all coated samples.  The results are shown in
Table~\ref{table:results}.  The values for the coating loss angles
$\phi_{\mathrm{coat,||}}$ are calculated from the modal Q's from
\begin{equation}
\label{eq:q} \phi_{\mathrm{coat, ||}} = \left( 1/Q_{\mathrm{coated}}
- 1/Q_{\mathrm{uncoated}}\right)/ \left( t \, \mathrm{d}U/U\right),
\end{equation}
where $Q$ is the modal Q, measured for the disk both coated and
uncoated, $t$ is the thickness of the coating, and $\mathrm{d}U/U$
is the ratio of energy stored in the coating per unit coating
thickness to the total energy for each given mode shape.  These
values of $\mathrm{d}U/U$ were calculated using a finite element
model\cite{Crooks}, using Young's moduli $Y_{\mathrm{Ta2O5}} = 1.4
\times 10^{11}$~Pa and $Y_{\mathrm{SiO2}} = 7.2 \times 10^{10}$~Pa,
and are shown in Table~\ref{table:ratios}. To determine the coating
loss coming from internal friction, the loss predicted from coating
thermoelastic damping\cite{Fejer,Braginsky03} was subtracted from
the $\phi$ calculated in Eq.~\ref{eq:q}.

\begin{table}
\caption{Results of mechanical loss measurements on TiO$_2$-doped
Ta$_2$O$_5$/SiO$_2$ coatings.  All coatings were 30 layers of
alternating material with various concentrations of TiO$_2$ in the
Ta$_2$O$_5$.  Each had an optical thickness  $\lambda/4$ in each
layer except for coating 6.  This coating was a single layer of
TiO$_2$-doped Ta$_2$O$_5$ 4.730~$\mu$m thick. All coatings were done
in the small coating chamber except where noted.}
\label{table:results}
\begin{center}
\begin{tabular}{llrrc}
\hline
Coating & Thickness &  Frequency & Modal Q & Loss angle $\phi_{||}$ \\
\hline
0 & Thin & 2733 & 5.4 $\times 10^{5}$ & 2.5 $\times 10^{-4}$ \\
  &      & 2735 & 5.3 $\times 10^{5}$ & 2.5 $\times 10^{-4}$ \\
  &      & 4130 & 4.3 $\times 10^{5}$ & 2.8 $\times 10^{-4}$ \\
  & Thick& 20180& 4.8 $\times 10^{6}$ & 2.2 $\times 10^{-4}$ \\
  &      & 20183& 3.6 $\times 10^{6}$ & 3.2 $\times 10^{-4}$ \\
  &      & 28383& 3.0 $\times 10^{6}$ & 3.3 $\times 10^{-4}$ \\
  &      & 28387& 3.3 $\times 10^{6}$ & 2.9 $\times 10^{-4}$ \\
  &      & 47349& 5.6 $\times 10^{6}$ & 3.7 $\times 10^{-4}$ \\
  &      & 47363& 6.3 $\times 10^{6}$ & 2.9 $\times 10^{-4}$ \\
  &      & 73454& 3.0 $\times 10^{6}$ & 4.0 $\times 10^{-4}$ \\
  &      & 73458& 3.8 $\times 10^{6}$ & 2.7 $\times 10^{-4}$ \\
1 & Thin & 2653 & 7.6 $\times 10^{5}$ & 1.8 $\times 10^{-4}$ \\
  &      & 2666 & 2.0 $\times 10^{5}$ & 7.1 $\times 10^{-4}$ \\
  &      & 4026 & 3.6 $\times 10^{5}$ & 3.6 $\times 10^{-4}$ \\
  &      & 6045 & 7.4 $\times 10^{5}$ & 1.9 $\times 10^{-4}$ \\
  &      & 6078 & 7.6 $\times 10^{5}$ & 1.8 $\times 10^{-4}$ \\
  & Thick& 20191& 4.9 $\times 10^{6}$ & 2.2 $\times 10^{-4}$ \\
  &      & 28428& 3.5 $\times 10^{6}$ & 2.6 $\times 10^{-4}$ \\
  &      & 47423& 5.4 $\times 10^{6}$ & 4.1 $\times 10^{-4}$ \\
  &      & 73515& 3.4 $\times 10^{6}$ & 3.4 $\times 10^{-4}$ \\
2 & Thin & 2706 & 8.9 $\times 10^{5}$ & 1.5 $\times 10^{-4}$ \\
  &      & 2711 & 8.8 $\times 10^{5}$ & 1.5 $\times 10^{-4}$ \\
  &      & 4101 & 6.8 $\times 10^{5}$ & 1.7 $\times 10^{-4}$ \\
  &      & 6165 & 8.4 $\times 10^{5}$ & 1.6 $\times 10^{-4}$ \\
  &      & 6184 & 8.6 $\times 10^{5}$ & 1.6 $\times 10^{-4}$ \\
  &      & 9464 & 6.3 $\times 10^{5}$ & 2.0 $\times 10^{-4}$ \\
  &      & 9465 & 6.1 $\times 10^{5}$ & 2.0 $\times 10^{-4}$ \\
  & Thick& 20239& 6.7 $\times 10^{6}$ & 1.9 $\times 10^{-4}$ \\
  &      & 28488& 5.4 $\times 10^{6}$ & 1.8 $\times 10^{-4}$ \\
  &      & 47466& 10.0 $\times 10^{6}$& 2.5 $\times 10^{-4}$ \\
  &      & 73599& 5.2 $\times 10^{6}$ & 2.5 $\times 10^{-4}$ \\
3 & Thin & 2722 & 9.3 $\times 10^{5}$ & 1.5 $\times 10^{-4}$ \\
  &      & 4111 & 6.4 $\times 10^{5}$ & 1.9 $\times 10^{-4}$ \\
  &      & 6197 & 9.1 $\times 10^{5}$ & 1.5 $\times 10^{-4}$ \\
  &      & 9517 & 6.4 $\times 10^{5}$ & 2.0 $\times 10^{-4}$ \\
  &      & 9519 & 6.6 $\times 10^{5}$ & 1.9 $\times 10^{-4}$ \\
  & Thick& 20245& 7.0 $\times 10^{6}$ & 1.7 $\times 10^{-4}$ \\
  &      & 28500& 6.3 $\times 10^{6}$ & 1.4 $\times 10^{-4}$ \\
  &      & 47485& 11.3 $\times 10^{6}$& 2.0 $\times 10^{-4}$ \\
  &      & 73620& 7.0 $\times 10^{6}$ & 1.5 $\times 10^{-4}$ \\
4$^{*}$ & Thin& 2723 & 9.2 $\times 10^{5}$ & 1.5 $\times 10^{-4}$ \\
  &      & 2724 & 9.7 $\times 10^{5}$ & 1.4 $\times 10^{-4}$ \\
  &      & 4114 & 6.0 $\times 10^{5}$ & 2.0 $\times 10^{-4}$ \\
  &      & 6200 & 8.1 $\times 10^{5}$ & 1.7 $\times 10^{-4}$ \\
  &      & 9524 & 6.1 $\times 10^{5}$ & 2.1 $\times 10^{-4}$ \\
  & Thick& 20241& 5.4 $\times 10^{6}$ & 2.5 $\times 10^{-4}$ \\
  &      & 28493& 4.1 $\times 10^{6}$ & 2.7 $\times 10^{-4}$ \\
  &      & 47467& 6.1 $\times 10^{6}$ & 5.3 $\times 10^{-4}$ \\
  &      & 73598& 5.0 $\times 10^{6}$ & 2.7 $\times 10^{-4}$ \\
5$^{*}$ & Thick& 20193&  6.5 $\times 10^{6}$& 2.0 $\times 10^{-4}$ \\
  &      & 28400& 5.4 $\times 10^{6}$ & 1.8 $\times 10^{-4}$ \\
  &      & 47398& 10.9 $\times 10^{6}$& 2.2 $\times 10^{-4}$ \\
  &      & 73524& 6.2 $\times 10^{6}$ & 1.9 $\times 10^{-4}$ \\
6$^{**}$ & Thick & 20226 & 2.9 $\times 10^{6}$& 2.6 $\times 10^{-4}$ \\
  &       & 28462 & 1.9 $\times 10^{6}$& 3.7 $\times 10^{-4}$ \\
  &       & 47430 & 3.6 $\times 10^{6}$& 4.4 $\times 10^{-4}$ \\
  &       & 73588 & 2.4 $\times 10^{6}$& 3.3 $\times 10^{-4}$ \\
\hline \multicolumn{4}{l}{$^{*}$Coated in large coating chamber}\\
\multicolumn{4}{l}{$^{**}$Single layer of TiO$_2$-doped Ta$_2$O$_5$
}
\end{tabular}
\end{center}
\end{table}

\begin{table}
\caption{Ratio of energy in the coating (per unit coating thickness)
to total energy for modes of the thin and thick samples.  }
\label{table:ratios}
\begin{center}
\begin{tabular}{lrr}
\hline
Thickness & Approximate modal frequency (Hz) & d$U/U$ (1/m) \\
\hline
Thin  &  2700   & 1584 \\
      &  4100   & 1659 \\
      &  6200   & 1581 \\
      &  9500   & 1624 \\
Thick & 20200   & 142.4 \\
      & 28500   & 153.2 \\
      & 47400   & 52.33 \\
      & 73500   & 109.1 \\
\hline
\end{tabular}
\end{center}
\end{table}

\begin{figure}
   \begin{center}
   \includegraphics[height=6 cm,angle=0]{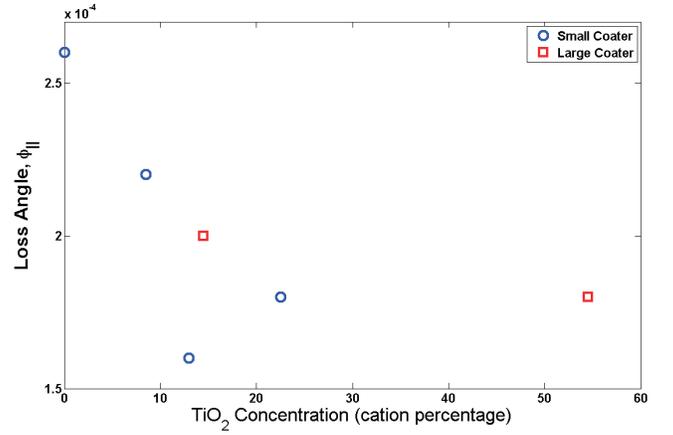}
   \end{center}
   \caption{Loss angle of TiO$_2$-doped Ta$_2$O$_5$/SiO$_2$ coating as a
function of TiO$_2$ concentration in the Ta$_2$O$_5$.  The TiO$_2$
concentration used was the one determined by EELS when available,
otherwise the concentration determined by interpolating index
changes. \label{fig:result}}
   \end{figure}

The coating loss for each concentration of TiO$_2$, calculated mode
by mode for all samples measured, was fit to a frequency dependant
model, following a recent paper\cite{Crooks2}.  The results are
shown in Table~\ref{table:freq}.  The results of extrapolating these
formulas to 100~Hz, where interferometers will be limited by coating
thermal noise, are shown in Fig.~\ref{fig:result}.

\subsection{Thermal noise}
\begin{figure}
   \begin{center}
   \includegraphics[height=5.5 cm,angle=0]{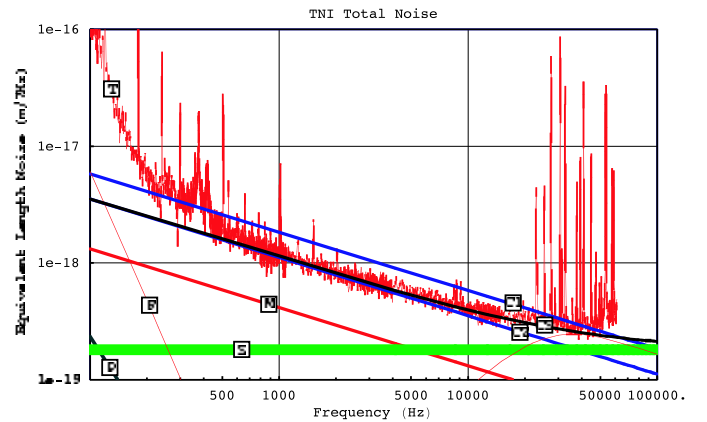}
   \end{center}
   \caption{Direct thermal noise measurement of a
   TiO$_2$-doped Ta$_2$O$_5$/SiO$_2$ coating.  The curve labeled ``T'' is
   the total noise spectrum of the interferometer. ``C1'' is the thermal
   noise of an undoped coating, as previously measured.
   ``C2'' is the thermal noise of the doped coating, where the loss
angle was adjusted to fit the data. ``S'' is the shot noise of the
instrument, and ``CS'' is the quadrature sum of this shot noise and
the doped-coating thermal noise. ``M'' is an upper bound on the
mirror, or substrate thermal noise, based on in-situ measurements of
mechanical Q's of the mirrors. ``F'' is the laser frequency noise,
and ``P'' is an upper bound on the pendulum thermal
noise.\label{fig:tni}}
   \end{figure}

A direct, interferometric broadband measurement was made of the
thermal noise of the TiO$_2$-doped Ta$_2$O$_5$/SiO$_2$ coatings
using the small scale interferometer at Caltech. The measurement
apparatus and the results for undoped coatings are described
elsewhere~\cite{tni}. Figure~\ref{fig:tni} shows the result for
coated mirrors done in a separate coating run, but using the same
coating formula as coating 2 in Table~\ref{table:results} except
that these mirrors were coated in LMA/Virgo's large coating chamber,
while the samples used for the Q measurements were done in the small
chamber. The value for the loss angle obtained by this direct
measurement and fit is
\begin{equation}
\phi_{\mathrm{eff}} = (2.41 \pm 0.15) \times 10^{-6},
\label{eq:tniresult}
\end{equation}
where $\phi_{\mathrm{eff}}$ is from Eq.~\ref{eq:phi}. The clear
reduction in thermal noise is shown graphically in
Fig.~\ref{fig:tni} and quantitatively between Eqs.~\ref{eq:tni} and
\ref{eq:tniresult}. The result in Eq.~\ref{eq:tniresult} is to be
compared with the value predicted from the modal Q results in
Table~\ref{table:freq},
\begin{equation}
\phi_{\mathrm{eff}} = (4.0 \pm 0.3) \times 10^{-6},
\label{eq:tniprediction}
\end{equation}
assuming $\phi_{\mathrm{SiO2}} = 1.0 \times 10^{-4}$.  The reason
for the discrepancy between Eqs.~\ref{eq:tniresult} and
\ref{eq:tniprediction} is not known.

\begin{table}
\caption{Coating mechanical loss fit to a frequency dependant model
for TiO$_2$-doped Ta$_2$O$_5$/SiO$_2$ coatings.  All coatings were
done in the small coating chamber except where noted.}
\label{table:freq}
\begin{center}
\begin{tabular}{lc}
\hline
Coating & $\phi_{\mathrm{coat, ||}}$ \\
\hline 0 & $\left(2.6 \pm 0.2 \right) \times 10^{-4} + f\left(1.2
\pm 0.6 \right) \times 10^{-9}$ \\
1 & $\left(2.2 \pm 0.4 \right) \times 10^{-4} + f\left(2.2
\pm 1.1 \right) \times 10^{-9}$ \\
2 & $\left(1.6 \pm 0.1 \right) \times 10^{-4} + f\left(1.4
\pm 0.3 \right) \times 10^{-9}$ \\
3 & $\left( 1.8 \pm 0.1 \right) \times 10^{-4} + f\left(-0.2
\pm 0.4 \right) \times 10^{-9}$  \\
4$^{*}$ & $\left(1.8 \pm 0.2 \right) \times 10^{-4} + f\left(1.7
\pm 0.6 \right) \times 10^{-9}$ \\
5$^{*}$ & $\left(2.0 \pm 0.2 \right) \times 10^{-4} + f\left(0.1
\pm 0.4 \right) \times 10^{-9}$ \\
6$^{**}$ & $\left(3.1 \pm 1.0 \right) \times 10^{-4} + f\left(0.1
\pm 2.1\right) \times 10^{-9}$\\
\hline \multicolumn{2}{l}{$^{*}$Coated in large coating chamber} \\
\multicolumn{2}{l}{$^{**}$Single layer of TiO$_2$-doped Ta$_2$O$_5$}
\end{tabular}
\end{center}
\end{table}

\subsection{Optical absorption}
Optical absorption was measured using photothermal, common-path
interferometry (PCI), which is a modified thermal lensing technique
that exploits the thermo-optic effect (index of refraction
dependence on temperature: d$n/\mathrm{d}T$).  It differs from
standard far-field thermal lensing by utilizing a near-field
detection scheme which approaches in sensitivity that of
interferometric absorption measurement methods. Phase distortions,
$\delta \phi$, of the probe beam due to heating by an intersecting
pump beam in a skewed cylindrical region (approximately 75 $\mu$m
diameter by x 500 $\mu$m long) are transformed into perturbations of
the probe beam intensity, $\Delta I/I \approx \Delta \phi$, that are
easily detectable using a lock-in detection technique which gives
both amplitude and phase. For materials with d$n/\mathrm{d}T$ around
$10^{-5}/$K and with pump powers of 1~W, resolutions of $\leq 1$~ppm
in terms of the absorbed fraction of pump power are readily
achievable. The signal phase can be used to discriminate between
probe light scattered from surface imperfections compared to that
diffracted by the thermal wave emanating from the heated surface.

\begin{figure}
   \begin{center}
   \includegraphics[height=6 cm,angle=0]{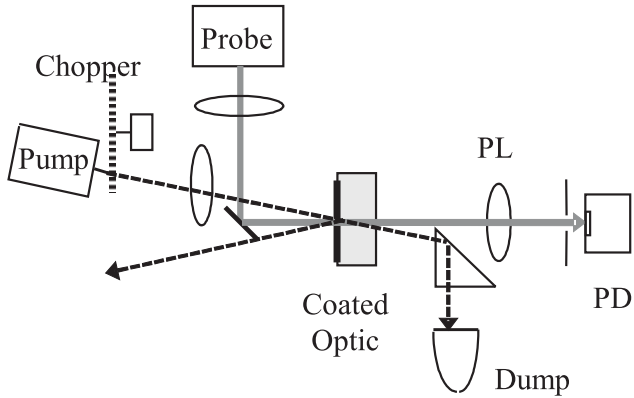}
   \end{center}
   \caption{. Crossed-beam setup for low absorption photo-thermal,
   common-path interferometry measurement. PL: projecting lens,
   PD: photodetector.
\label{fig:pci}}
   \end{figure}

The PCI technique used here utilizes a chopped pump beam at 1064~nm
which is crossed with a wider 632.8~nm probe beam inside the sample,
see Fig.~\ref{fig:pci}. In the case of high reflectivity (HR)
coatings, virtually all the pump beam is reflected. As long as
substrate losses are negligible as is the case with fused silica,
the heat deposited in the coating due to optical absorption of the
pump wavelength is the only source that heats the underlying
substrate (by thermal conduction).  In order for the probe beam to
sense the change in the local optical index via the resulting phase
distortions, it is necessary that the multi-layer dielectric HR
coatings be transparent to the probe beam, as are the coatings
studied here.  Results of optical absorption measurements as well as
index of refraction results are shown in Table~\ref{table:abs}.

\begin{table} \caption{Optical absorption of TiO$_2$-doped
Ta$_2$O$_5$/SiO$_2$ coatings and indices of refraction of individual
TiO$_2$-doped Ta$_2$O$_5$ layers within those coatings. Index of
refraction comparisons are only valid between coatings from the same
coating chamber.} \label{table:abs}
\begin{center}
\begin{tabular}{lcc}
\hline
Coating & Index $n$ & Absorption (ppm) \\
\hline
0 &  $2.065 \pm 0.005$ &  $0.9 \pm 0.2$ \\
1 &  $2.075 \pm 0.005$ &  $1.1 \pm 0.1$ \\
2 &  $2.092 \pm 0.005$ &  $1.0 \pm 0.1$ \\
3 &  $2.119 \pm 0.005$ &  $1.1 \pm 0.1$ \\
4$^{*}$ & $2.180 \pm 0.005$ & $2.5 \pm 0.5$ \\
5$^{*}$ & $2.070 \pm 0.005$ & $0.9 \pm 0.1$  \\
6$^{**}$& $2.075 \pm 0.005$ & $4.5 \pm 0.5$ \\
\hline \multicolumn{3}{l}{$^{*}$ Coated in large coating chamber} \\
\multicolumn{3}{l}{$^{**}$ Single layer of TiO$_2$-doped
Ta$_2$O$_5$}
\end{tabular}
\end{center}
\end{table}

In these experiments, a commercial neutral density (ND) filter
consisting of a partially transparent metallic film on a low-loss
fused silica substrate is used as a coating optical absorption
standard. In the PCI technique, the detected signal amplitude is
linear with respect to pump beam intensity and, therefore,
straightforward linear scaling can be used to relate the signal
levels from unknowns to those of the easily measured ND filters.
With both coated standards and unknowns, the coated surfaces are
positioned near the center of the sampling volume where maximum
signal is achieved. With 10~W of pump power, resolutions of $\leq
0.1$~ppm are achievable.

\section{Conclusion}
\label{sect:conc}

The mechanical loss results in Table~\ref{table:freq} and
Fig.~\ref{fig:result} show that adding TiO$_2$ to the Ta$_2$O$_5$
reduces the mechanical loss.  Differing concentrations of TiO$_2$ do
not effect the loss nearly as much as simply the presence or absence
of TiO$_2$.  This reduction of nearly half in the loss angle of the
coating corresponds to a significant improvement in thermal noise,
which translates into greater astronomical reach for advanced
interferometers. The optical absorption seen in
Table~\ref{table:abs} would be problematic in an advanced
interferometer, but slight improvements in coating technique may be
able to bring these numbers down to acceptable levels. The addition
of TiO$_2$ does appear to increase the optical absorption slightly,
so using minimal concentrations will be useful.  Changes in
annealing cycles are known to effect optical absorption, as does
levels of contamination, so both of these variables could
potentially be improved as well. Further measurements on the
inhomogeneity of the optical loss, the scatter, and the
reproducibility of all properties will be necessary before a
TiO$_2$-doped Ta$_2$O$_5$/SiO$_2$ could be accepted for use in an
actual gravitational-wave interferometer.

\section*{Acknowledgements}
\label{sect:ack} The LIGO Observatories were constructed by the
California Institute of Technology and Massachusetts Institute of
Technology with funding from the National Science Foundation under
cooperative agreement PHY-9210038.  The LIGO Laboratory operates
under cooperative agreement PHY-0107417.  We thank for Zhigang Pan
and  Slawomir Gras for checking the thermal noise formulas.  We also
gratefully acknowledge funding from NSF grant PHY-0355118(HWS). This
paper has been assigned LIGO Document Number LIGO-P050048-00-R.

\section*{Appendix: Measurement of the composition of the Ta$_2$O$_5$-TiO$_2$ layers}

\begin{figure}
   \begin{center}
   \includegraphics[height=6 cm,angle=0]{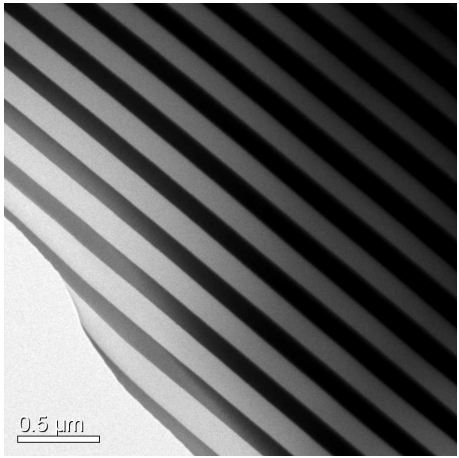}
   \end{center}
   \caption{Bright field TEM image of the multilayer structure:
   the lighter layers are the amorphous SiO$_2$ layers and the darker
   layers are the Ta$_2$O$_5$-TiO$_2$ layers. \label{fig:tem} }
   \end{figure}

The measurement of Ti dopant concentration was performed using
electron energy loss spectroscopy (EELS)~\cite{eels} on a FEI Tecnai
T20 transmission electron microscope operated at 200~kV equipped
with a Gatan Image Filter (GIF).  A bright field TEM image of a
coating is shown in Fig.~\ref{fig:tem}.  The edges used for
quantification were the O-K edge at 532~eV, the Ti-L$_{2,3}$ pair
starting at 456~eV, and the Ta-N$_{4,5}$ edge at 229~eV.  The O-K
and Ti-L edges arise from fairly simple atomic shells and
Hartree-Slater calculations of the partial cross sections are
readily available and reliable.  In contrast to this, N-shells are
rather more complex and no analytical representation of this partial
cross section was available.  This problem was circumvented by
determining the partial cross section experimentally from 20 spectra
from the Ta$_2$O$_5$ layers in the undoped coating 0, assuming that
the Ta:O ratio was the stoichiometric 2:5. This was then used with
the calculated Ti partial cross section to quantify spectra recorded
for the three coating samples 1, 3 and 4.

Quantification of the doped Ta$_2$O$_5$ layers in coating 1 was
carried out using 20 spectra collected from different thin regions
($< 50$~nm) and gave a result of $8.5 \pm 1.2$ cation percentage of
Ti (with the balance Ta). Only 4 spectra could be collected from
coating 3 due to much of the sample being too thick for accurate
EELS analysis, quantification yielded a result of $22.5 \pm 2.9$
cation percentage of Ti.  For coating 4, 20 spectra were recorded,
but more difficulties were encountered with the analysis. Accurate
quantification relies on correct subtraction of the background under
the edge, but this can become difficult when the edge is only a
small feature above the background.  The Ta edge is fairly weak and
has a delayed onset and background subtraction was a problem in
coating 4. Different background models were tried, some which
subtracted too much, and others which subtracted too little. The
best-fit background model led to a consistent quantification of the
20 spectra to give $54 \pm 5$ cation percentage of Ti, although it
is believed that too much background was removed here.  An
alternative background model which did not remove enough led to
quantifications in the low 40's (cation percentage of Ti).  It seems
likely that the real figure is about 5 cation percentage lower than
the above quoted figure and is of the order of 50 cation percentage
Ti.

\end{document}